# Communicability and Communities in Complex Socio-Economic Networks

**Ernesto Estrada and Naomichi Hatano**

Department of Mathematics, Department of Physics and Institute of Complex Systems, University of Strathclyde, Glasgow G11XQ

Institute of Industrial Science, University of Tokyo, Komaba, Meguro 153-8505, Japan

**Abstract**

The concept of communicability is introduced for complex socio-economic networks. The communicability function expresses how an impact propagates from one place to another in the network. This function is used to define unambiguously the concept of socio-economic community. The concept of temperature in complex socio-economic networks is also introduced as a way of accounting for the external stresses to which such systems are submitted. This external stress can change dramatically the structure of the communities in a network. We analyze here a trade network of countries exporting 'miscellaneous manufactures of metal.' We determine the community structure of this network showing that there are 27 communities with diverse degree of overlapping. When only communities with less than 80% of overlap are considered we found 5 communities which are well characterized in terms of geopolitical relationships. The analysis of external stress on these communities reveals that several countries are very much influenced by these critical situations, i.e., economical crisis. These weakest links are clearly identified and represent countries that are isolated from the main trade as soon as the external "temperature" of the system is increased. The current approach adds an important tool for the analysis of socio-economic networks in the real-world.

**Acknowledgments**
EE thanks B. Álvarez-Pereira and K. Deegan for help with the calculations. EE also thanks partial financial support from the New Professor's Fund given by the Principal, University of Strathclyde.

# 1 Introduction

There is no doubt that we live in a networked world. We live our lives connected by our family, friend and workmate ties. These social networks connect us to other people in other parts of the world to which we are only a few steps apart. In fact, we live in a "small-world." We also live surrounded by infrastructures which form their proper networks, such as supply networks, transportation networks, etc. The economic world in which we also live does not look very much different from this picture. Banks, corporations, industries and even universities and governmental institutions are interconnected forming a complex network of economical and political interrelations. Then, our lives are very much influenced by the structure of these complex networks and by the dynamics of processes taking place on them [1]. In this chapter we are interested in a tiny part of these complex worlds, namely in the analysis of the communicability and community structure of socio-economic networks.

There are different kinds of networks that can be analyzed in a socio-economic context. In general, economics is based on the activity of economic entities in business networks in which such entities are related by business relationships, such as the ones existing between companies and banks [2]. On a macroeconomical scale, world economies are becoming more and more interrelated as a consequence of the globalization, which implies control of capital flow and liberalization of trade policies [3]. Consequently, the study of the international trade network in which countries are represented by the nodes of the network and their commercial trades by the links is an important tool in understanding the interaction channels between countries [4].

The representation of socio-economic systems as networks gives us the possibility of analyzing theoretically how these interrelations are organized and how they influence the dynamics of processes taking place on them. For instance, network analysis permits to understand how economic perturbations in one country can spread to the whole world. In a complex network the nodes represent the entities of the system and the links their interactions [5]. Then, we consider socio-economic networks represented by simple graphs $G := (V, E)$. That is, graphs having $|V| = n$ nodes and $|E| = m$ links, without self-loops or multiple links between nodes. Let $\mathbf{A}(G) = \mathbf{A}$ be the adjacency matrix of the graph whose elements $A_{ij}$ are ones or zeroes if the corresponding nodes $i$ and $j$ are adjacent or not, respectively. In the next sections we introduce the concept of communicability between the entities in a complex socio-economic network and then analyze how to determine the community structure of such networks.

## 2 The concept of communicability

It is known that in many situations the communication between a pair of nodes in a network does not take place only through the optimal route connecting both nodes. In such situations, the information can flow from one node to another by following other non-optimal routes. Here we consider that a pair of nodes has good *communicability* if there are several non-optimal routes connecting them in addition to the optimal one. The optimal route can be identified in the case of a simple network with the shortest path connecting both nodes. A path is a sequence of different nodes and links in connecting both nodes. Let $s$ be the length of this shortest path. Then, the non-optimal routes correspond to all walks of lengths $k > s$ which connect both nodes. A walk of length $k$ is a sequence of (not necessarily different) vertices $v_0, v_1, \cdots, v_{k-1}, v_k$ such that for each $i = 1, 2 \cdots, k$ there is a link from $v_{i-1}$ to $v_i$. Consequently, these walks communicating two nodes in the network can revisit nodes and links several times along the way, which is sometimes called "backtracking walks." We assume that the shortest walks are more important than the longer ones [6].

> The communicability between a pair of nodes $p, q$ in the graph is defined as the weighted sum of all walks starting at node $p$ and ending at node $q$, giving more weight to the shortest walks.

It is well-known that the $(p, q)$-entry of the $k$ th power of the adjacency matrix, $(\mathbf{A}^k)_{pq}$, gives the number of walks of length $k$ starting at the node $p$ and ending at the node $q$ [7]. Then, the communicability function can be expressed by the following formula [6]:

$$G_{pq} = \sum_{k=0}^{\infty} c_k (\mathbf{A}^k)_{pq}. \tag{1}$$

The coefficients $c_k$ need to fulfill the following requirements: (i) make the series (1) converges, (ii) giving less weight to longer walks, and (iii) giving real positive values for the communicability. For the sake of simplicity we select here $c_k = 1/k!$ [6], which gives the following communicability function:

$$G_{pq} = \sum_{k=0}^{\infty} \frac{(\mathbf{A}^k)_{pq}}{k!} = (e^{\mathbf{A}})_{pq}. \tag{2}$$

The right-hand side (RHS) of the expression (2) corresponds to the non-diagonal entry of the exponential adjacency matrix. Let $\lambda_1 \geq \lambda_2 \geq \cdots \geq \lambda_n$ be the eigenvalues of the adjacency matrix in the non-increasing order and let $\varphi_j(p)$ be the $p$ th entry of the $j$ th eigenvector which is associated with the eigenvalue $\lambda_j$ [7]. Then, using the spectral decomposition of the adjacency matrix the communicability function can be written as [6]

$$G_{pq} = \sum_{j+1}^{n} \varphi_j(p)\varphi_j(q)e^{\lambda_j}. \qquad (3)$$

In order to understand the communicability function in the context of complex socio-economic networks let us consider the following examples. First, let us consider a series of corporations having business relationships in such a way that they form a linear chain as the one depicted below:

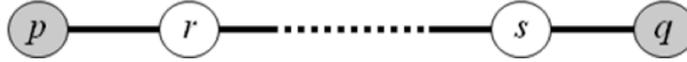

The communicability between the corporations $p$ and $q$, which are the endpoints of the chain is given by the following mathematical expression

$$G_{pq} = \frac{1}{n+1} \sum_j \left( \cos\frac{j\pi(p-q)}{n+1} - \cos\frac{j\pi(p+q)}{n+1} \right) e^{2\cos\left(\frac{j\pi}{n+1}\right)}, \qquad (4)$$

where we have used $p$ and $q$ to designate the number of these nodes in the chain starting by 1. For instance, if the chain is formed by only 4 companies then $p = 1$ and $q = 4$. In this case, the communicability between the corporations $p$ and $q$ tends to zero when the number of corporations is very large. That is, $G_{pq} \to 0$ when $n \to \infty$.

A completely different picture emerges when we consider that all corporations are interrelated to each other forming a compact cluster as the one illustrated below:

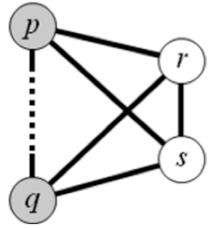

In this case it is easy to show that the communicability between any pair of nodes is given by the following expression:

$$G_{pq} = \frac{1}{ne}(e^n - 1). \tag{5}$$

Consequently, when the number of corporations is very large the communicability between any pair of them tends to infinity, i.e., $G_{pq} \to \infty$ when $n \to \infty$.

These results indicate that a simple linear interdependence between the corporations gives rise to very poor communicability between the entities involved. On the other hand, a fully interconnected network of corporations is quite effective for the communicability between the entities involved. However, this kind of fully interconnected organizations is very rare at large scales due to infrastructural and cost-benefit reasons. In general, socio-economic networks display complex organizational structures which are neither linear nor fully connected. Then, the communicability function allows us to study the patterns of communication hidden in such structures.

## 3 Communicability and Socio-Economic Communities

A community in a complex socio-economic network can be understood as a set of entities displaying large internal cohesion. This means that the members of the community are more "tightly connected" among them than with the outsiders of the community. The concept of communicability introduces an intuitive way for finding the structure of communities in complex networks. In this context a socio-economic community is a subset of entities that have larger communicability among the members of the community than with the rest of the entities in the network.

In order to define a socio-economic community we first carry out an analysis of the communicability function (3). The term $\varphi_j(p)\varphi_j(q)e^{\lambda_j}$ can be positive or negative on the basis of the signs of the $p$ th and $q$ th components of the corre-sponding eigenvector. A better understanding of this sign pattern can be obtained using the following physical analogy. Suppose that the nodes of the network in question are balls and the links are springs. We argued [6] that the adjacency matrix $\mathbf{A}$ is equivalent to the Hamiltonian of the spring network and the communicability (3) is equivalent to the Green's function of the spring network. Then, the eigenvectors of the adjacency matrix represent vibrational normal modes of the network. The sign of the $p$ th component of the $j$ th eigenvector indicates the direction of the vibration, say up or down. If two nodes, $p$ and $q$, have the sign for the $j$ th eigenvector it indicates that these two nodes are vibrating in the same direction. In Fig. 1 we illustrate the directions of the vibrational modes for the nodes in a simple network. According to the Perron-Frobenius theorem [7] all the components of the principal eigenvector $\varphi_1$ has the same sign. Consequently, it represents the translational movement of the whole network.

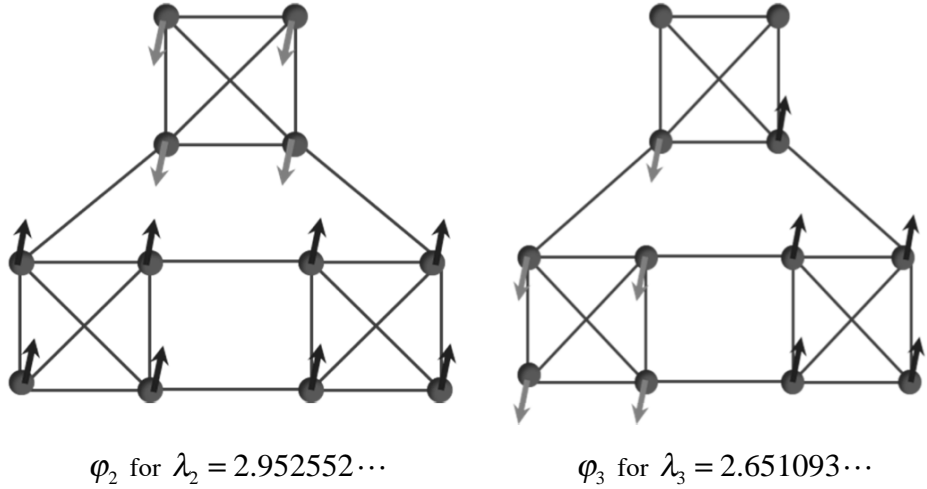

$\varphi_2$ for $\lambda_2 = 2.952552\cdots$ $\qquad\qquad$ $\varphi_3$ for $\lambda_3 = 2.651093\cdots$

**Fig. 1.** Illustration of the vibrational modes of the nodes in a network corresponding to the second and third largest eigenvalue.

Using this physical analogy we can decompose the communicability function (3) as follows [6]:

$$G_{pq} = \left[\varphi_1(p)\varphi_1(q)e^{\lambda_1}\right]$$
$$+ \left[\sum_{2 \leq j \leq n}{}^{++}\varphi_j(p)\varphi_j(q)e^{\lambda_j} + \sum_{2 \leq j \leq n}{}^{--}\varphi_j(p)\varphi_j(q)e^{\lambda_j}\right] \quad (6)$$
$$+ \left[\sum_{2 \leq j \leq n}{}^{+-}\varphi_j(p)\varphi_j(q)e^{\lambda_j} + \sum_{2 \leq j \leq n}{}^{-+}\varphi_j(p)\varphi_j(q)e^{\lambda_j}\right].$$

The first bracketed term on the right-hand side of the Green's function (6) represents the *movement* of all the nodes (the balls) in one direction after an impact on one node, as if they were part of a giant cluster formed by the whole. In the second bracketed term on the right-hand side of Eq. (6), the nodes $p$ and $q$ have the same sign of the corresponding eigenvector (positive or negative); if we put an impact on the ball $p$, the ball $q$ oscillates in the same direction as the ball $p$. We thus regard that $p$ and $q$ are in the same cluster if there are more than one cluster in the network. Consequently, we call this second term of Eq. (6) the *intracluster communicability*. The last bracketed term of Eq. (6), on the other hand, represents an uncoordinated movement of the nodes $p$ and $q$, i.e., they have different signs of the eigenvector component; if we put an impact on the ball $p$, the ball $q$ oscillates in the opposite direction. We regard that they are in different clusters of the network. Then, we call this third term of Eq. (6) the *intercluster communicability* between a pair of nodes.

As we are interested in the community structure of the network which is determined by the clustering of nodes into groups, we leave out the first term from Eq. (6) because we are not interested in the translational movement of the whole network and thereby consider the quantity

$$\Delta G_{pq} = \left[\sum_{2 \leq j \leq n}{}^{++}\varphi_j(p)\varphi_j(q)e^{\lambda_j} + \sum_{2 \leq j \leq n}{}^{--}\varphi_j(p)\varphi_j(q)e^{\lambda_j}\right]$$
$$+ \left[\sum_{2 \leq j \leq n}{}^{+-}\varphi_j(p)\varphi_j(q)e^{\lambda_j} + \sum_{2 \leq j \leq n}{}^{-+}\varphi_j(p)\varphi_j(q)e^{\lambda_j}\right] \quad (7)$$
$$= \sum_{j=2}^{\text{intracluster}}\varphi_j(p)\varphi_j(q)e^{\lambda_j} - \left|\sum_{j=2}^{\text{intercluster}}\varphi_j(p)\varphi_j(q)e^{\lambda_j}\right|,$$

where in the last line we used the fact that the intracluster communicability is positive and the intercluster communicability is negative [6]. Now, we are in conditions of defining a socio-economic community in an unambiguous way.

> A socio-economic community is a group of entities $C \subseteq V$ in the network $G = (V, E)$ for which the intracluster communicability is larger than the intercluster one, $\Delta G_{p,q}(\beta) > 0 \quad \forall (p, q) \in C$.

Using this definition we can generate algorithms that identify the communities in a network without the use of any external parameter. This is the topic of the next section.

## 3.1 Communicability graph

In order to find communities in a complex network we need to identify all pairs of nodes having $\Delta G_{p,q}(\beta) > 0$ in the network. Let us start by representing the values of $\Delta G_{p,q}(\beta)$ as the non-diagonal entries of the matrix $\Delta(G)$, for which the diagonal entries are zeroes. We are interested only the positive entries of this matrix, which correspond to the pairs of nodes having larger intra- than intercluster communicability. Then, let us introduce the following Heavyside function,

$$\Theta(x) = \begin{cases} 1 & \text{if } x \geq 0 \\ 0 & \text{if } x < 0. \end{cases} \tag{8}$$

If we apply this function in an elementwise way to the $\Delta(G)$ matrix we obtain a symmetric binary matrix having ones for those pairs of nodes having $\Delta G_{p,q}(\beta) > 0$ and zero otherwise. This matrix can be represented as a new graph, which we call the *communicability graph* $\Theta(G)$. The nodes of $\Theta(G)$ are the same as the nodes of $G$, and two nodes $p$ and $q$ in $\Theta(G)$ are connected if, and only if, $\Delta G_{p,q}(\beta) > 0$ in $G$. Now, a community can be identified as a locally maximal complete subgraph in the network. We recall that a *complete subgraph* is a part of a graph in which all nodes are connected to each other. If this complete subgraph is maximal it is known as a *clique*. The following immediately gives us the method for identifying communities in a complex network [8].

> A community is a clique in the communicability graph.

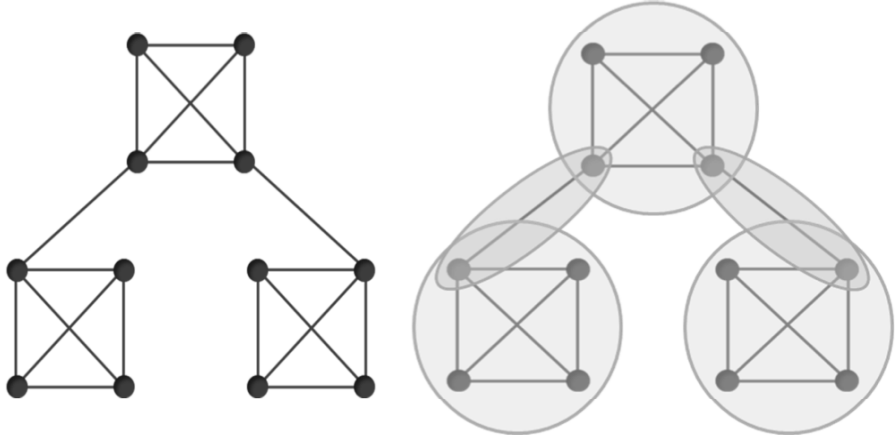

**Fig. 2.** The communicability graph (left) of the network displayed in Fig. 1 and the cliques existing in this communicability graph, which correspond to the communities of the network in Fig. 1.

In Fig. 2 we illustrate the communicability graph for the simple network displayed in Fig.. 1. There are three 4-nodes cliques and two 2-nodes cliques in this communicability graph, which are also illustrated in this figure.

## 3.2 Overlapping communities

As can be seen in Fig. 2 there are some nodes that are in more than one community at the same time. Then, the corresponding communities overlap with each other in certain degree. Here we deal with the overlapping of communities and how to merge them to form larger communities.

Two communities are overlapped if they share at least one common node. We can use this information in order to analyze the degree of overlapping between two communities, which can be related to the similarity between the communities in question. Then, we propose the following index as the overlap between the communities *A* and *B* in a network [8]:

$$S_{AB} = \frac{2|A \cap B|}{|A| + |B|}, \qquad (9)$$

where the numerator is the number of nodes in common in the two communities and the denominator gives the sum of the number of nodes in both communities. This index is known in the statistical literature as the Sørensen similarity index [10] and is used to compare the similarity between two samples in ecological sys-tems in particular. The index is bounded as $0 \leq S_{AB} \leq 1$, where the lower bound is obtained when no overlap exists between the two communities and the maximum is reached when the two communities are identical. We can calculate the similarity index $S_{AB}$ for each pair of communities found in the network and represent all results as an overlapping matrix $\mathbf{S}$. If we are interested in identifying only those communities that have an overlap lower than a certain value $\alpha$, $S_{AB} < \alpha$, we merge those communities for which $S_{AB} \geq \alpha$ into simpler communities.

We have proposed the following general algorithm for the mergence of communities in a complex network [8]:

1. Find the communities in the network following the approach described in the preceding sections;
2. Calculate $S_{AB}$ for all pairs of communities found in the previous step and build the matrix $\mathbf{S}$;
3. For a given value of $\alpha$, build the matrix $\mathbf{O}$, whose entries are given by

$$O_{AB} = \begin{cases} 1 & \text{if } S_{AB} \geq \alpha, \\ 0 & \text{if } S_{AB} < \alpha, \text{ or } A = B; \end{cases}$$

4. If $\mathbf{O} = \mathbf{0}$, go to the end; else go to the step (v);
5. Enumerate the cliques in the graph whose adjacency matrix is $\mathbf{O}$. Every clique in $\mathbf{O}$ represents a group of communities with overlaps larger than or equal to $\alpha$;
6. Build the merged communities by merging the communities represented by the nodes forming the cliques found in the step (iv) and go to the step (ii);
7. End.

## 4 Socio-economic communities under external "stress"

Complex networks in general and their communities in particular are exposed to external "stress," which are independent of the organizational architecture of the network. For instance, the web of social relations between actors in a society depends on the level of "social agitation" existing in such society in the specific period of time under study. The network of economic relations between industries and/or banks is affected by the existence of economical crisis, which can change the architecture of such interdependences. The challenge is to capture these external stresses into a quantity that allows us to model the evolution of communities in a complex socio-economic network.

We have proposed to consider the following physical analogy to study this problem [10]. Let us consider that the complex network is submerged into a thermal bath at the temperature $T$. The thermal bath represents the external situation which affects all the links in the network at the same time. Then, after equilibration all links in the network will be weighted by the parameter $\beta = (k_B T)^{-1}$. The parameter $\beta$ is known as the *inverse temperature*.

Then, the communicability between a pair of nodes in a network at the inverse temperature $\beta$ is given by

$$G_{p,q}(\beta) = \sum_{j=1}^{n} \phi_j(p)\phi_j(q)e^{\beta \lambda_j} . \qquad (10)$$

The expression (10) tells us that the structure of the communities depends on the external stress at which the network is submitted. For instance, as $\beta \to 0$ ($T \to \infty$), the communicability between any pair of nodes in the graph vanishes as

$$G_{pq}(\beta \to 0) = \sum_{j=1}^{n} \phi_j(p)\phi_j(q) = 0 . \qquad (11)$$

In other words, when the external stress is very large, like in a crisis situation there is no communicability between any pair of entities in the network. This situation resembles the case where any individual or corporation is disconnected from the rest in a globally individualistic behaviour.

On the other extreme as $\beta \to \infty$ ($T \to 0$), the communicability between any pair of nodes in the graph is determined by the principal eigenvalue/eigenvector of the adjacency matrix,

$$G_{pq}(\beta \to \infty) = \phi_1(p)\phi_1(q)e^{\lambda_1} , \qquad (12)$$

which means that all entities in the complex network form a unique community. This situation resembles the case where there is an ideal stability in the system which allows all entities to be interrelated to each other forming a unique club in the network. In Fig. 3 we represent these extreme situations for the network represented in Fig. 1.

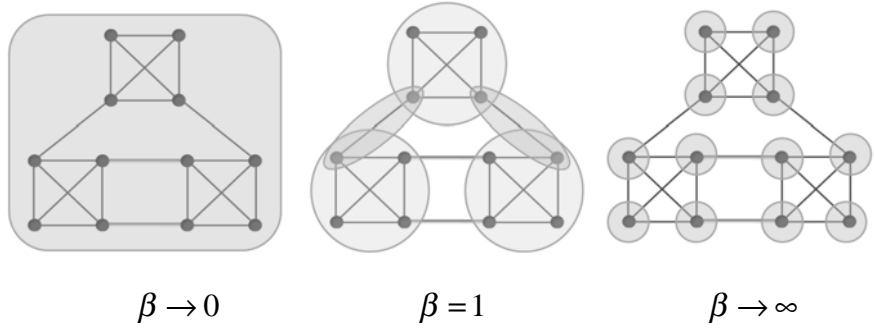

$\beta \to 0$     $\beta = 1$     $\beta \to \infty$

**Fig. 3.** Illustration of the effect of external stress, accounted for by the inverse temperature, on the structure of communities in a complex network.

## 5 A social network illustration

As an illustration of the concepts explained in the previous sections we consider a friendship network known as the Zachary karate club, which has 34 members (nodes) with some friendship relations (links) [11]. The members of the club, after some entanglement, were eventually fractioned into two groups, one formed by the followers of the instructor and the other formed by the followers of the administrator. This network is illustrated in Fig. 4 in which the nodes are divided into the two classes observed experimentally by Zachary on the basis of the friendship relationships among the members of the club.

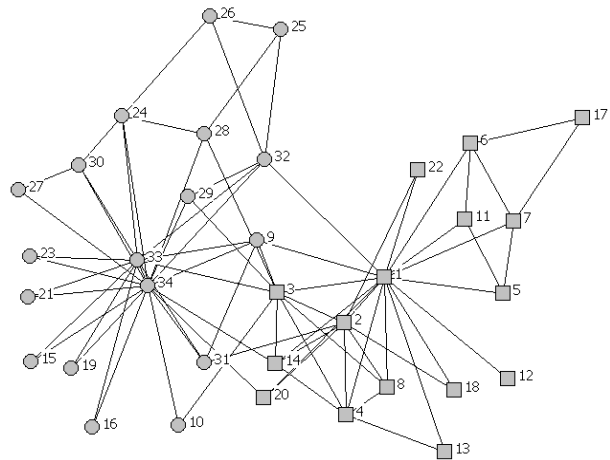

**Fig. 4.** Network representation of the Zachary karate club with the nodes represented as squares and circles according to the experimental classification made by Zachary.

The first step in indentifying the communities in this network is to obtain the communicability graph. In doing so we need to calculate the matrix $\Delta(G)$ and then dichotomize it using the function $\Theta(G)$. This matrix is the adjacency matrix of the communicability graph, which for the Zachary karate club is illustrated in Fig. 5.

By finding the cliques in this communicability graph, we detected the following communities [8]:

$A$ : {10,15,16,19,21,23,24,26,27,28,29,30,31,32,33,34} ;
$B$ : {9,10,15,16,19,21,23,24,27,28,29,30,31,32,33,34} ;
$C$ : {10,15,16,19,21,23,24,25,26,27,28,29,30,32,33,34} ;
$D$ : {1,2,3,4,5,6,7,8,11,12,13,14,17,18,20,22} ;
$E$ : {3,10} .

These communities display a large overlapping between them as can be seen in Fig.. 6.

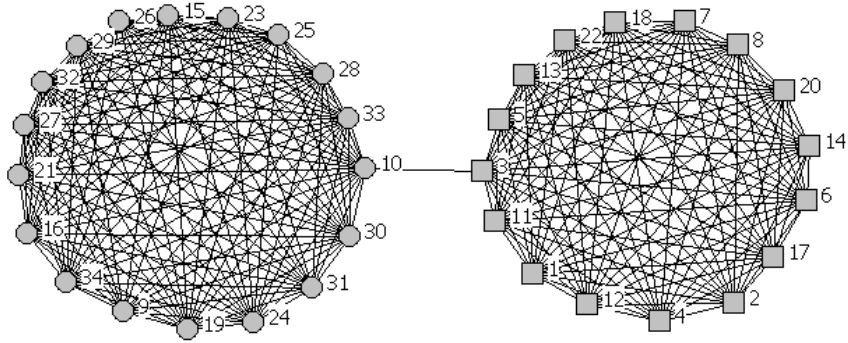

**Fig. 5.** Communicability graph corresponding to the Zachary karate club network.

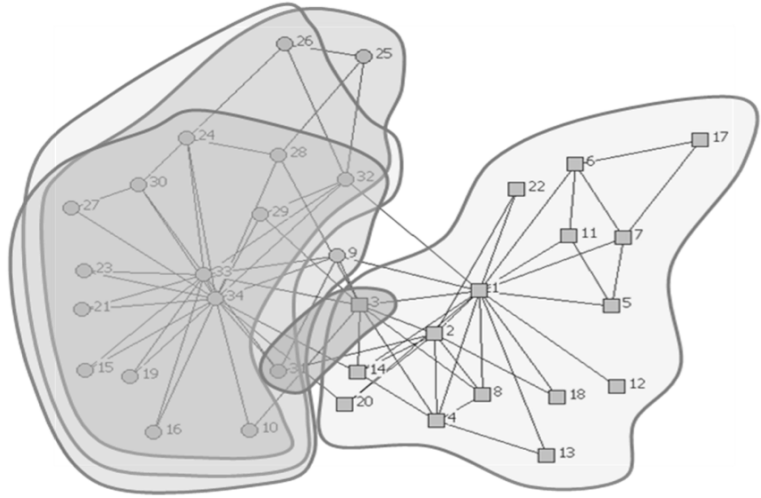

**Fig. 6.** Illustration of the overlapped communities found in the Zachary karate club network.

The large overlapping between these communities can be observed in the community-overlap matrix $\mathbf{S}$ for this network, which is given below:

$$\mathbf{S} = \begin{bmatrix} 1.000 & 0.938 & 0.938 & 0.000 & 0.111 \\ & 1.000 & 0.875 & 0.000 & 0.111 \\ & & 1.000 & 0.000 & 0.111 \\ & & & 1.000 & 0.111 \\ & & & & 1.000 \end{bmatrix}. \qquad (13)$$

If we consider only those communities having less than 10% of overlap we find only two communities in the network: $C_1 = A \cup B \cup C \cup E$ and $C_2 = D \cup E$. These two communities match perfectly the two factions found experimentally by Zachary in his study for the karate club. These two larger communities are illustrated in Fig. 7 (top-left graphic).

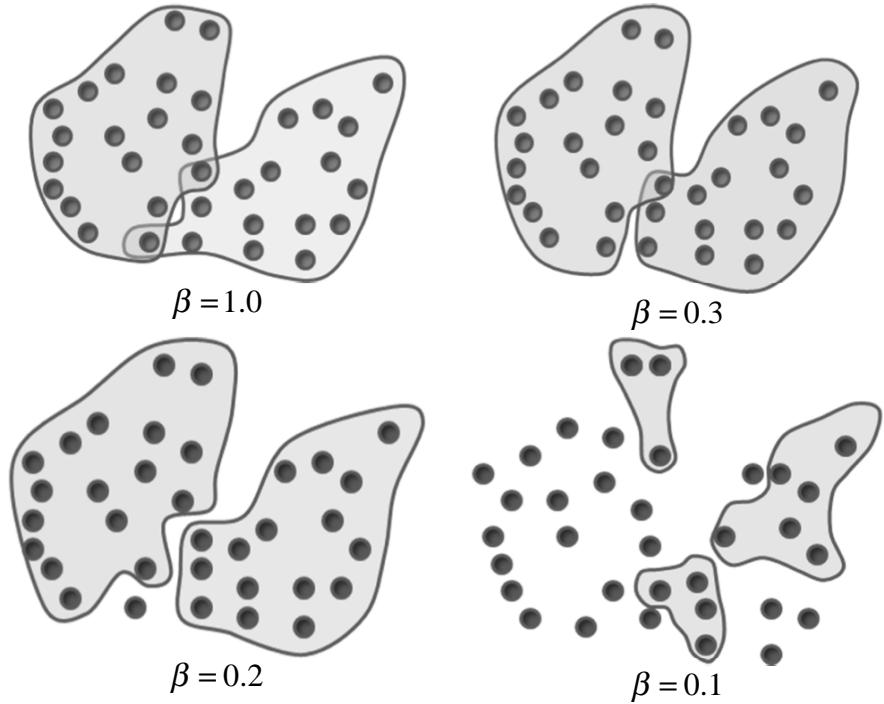

**Fig. 7.** Illustration of the effect of the temperature on the structure of communities in the Zachary karate club network. The links between individuals have been removed for the sake of simplicity.

In Fig. 7 we also illustrate the effect of the temperature on the structure of communities in the network. As the temperature increases the communities start to be fractioned until all nodes are forming individual communities. The case $\beta = 0$ represents a high level of stress, like a large social agitation. At this temperature the network structure is destroyed and every individual behaves independently. As the value of $\beta$ increases the stress at which the network is subjected decreases and several organizations of the society start to appear. In an ideal situation of no stress, $\beta \to \infty$, there is only one community in the network. Consequently, the consideration of the parameter $\beta$ permits to analyze the characteristics of the community structure of a network under different external conditions by considering that such conditions affect homogeneously to the nodes of the network.

## 6 Trade miscellaneous manufactures of metal

Here we study a real-world dataset on trade miscellaneous manufactures of metal among 80 countries in 1994. The data was compiled [12] for all countries with entries in the paper version of the Commodity Trade Statistics published by the United Nations. For some countries the authors used the 1993 data (Austria, Seychelles, Bangladesh, Croatia, and Barbados) or the 1995 data (South Africa and Ecuador) because they were not available for the year 1994. Countries which are not sovereign are excluded because additional economic data were not available: Faeroe Islands and Greenland, which belong to Denmark, and Macau (Portugal). Most missing countries are located in central Africa and the Middle East, or belong to the former USSR.

The network compiled by de Nooy [12] represents a weighted directed network. The arcs represent imports by one country from another for the class of commodities designated as "miscellaneous manufactures of metal" (MMM), which represents high technology products or heavy manufacture. The absolute value of imports (in 1,000 US$) is used but imports with values less than 1% of the country's total imports were omitted.

In general, all countries that export MMM also import them from other countries. However, there are 24 countries that are only importers and have no export of MMM at all. They are: Kuwait, Latvia, Philippines, French Guiana, Bangladesh, Fiji, Reunion, Madagascar, Seychelles, Martinique, Mauritius, Belize, Morocco, Sri Lanka, Algeria, Nicaragua, Iceland, Oman, Pakistan, Cyprus, Paraguay, Guadalupe, Uruguay, and Jordan. If the degree of the nodes is analyzed it is observed that the countries having the larger number of exports are (in order): Germany, USA, Italy, U.K., China, Japan, France, Belgium/Luxemburg, Netherland and Sweden. When an undirected version of this network is considered the same countries appear as the larger exporters, with only tiny variations in the order: Germany, USA, Italy, U.K., Japan, China, France, Netherland, Belgium/Luxemburg, Sweden. On the other hand, the in-degree is practically the same for every country. For instance, the average in-degree is 12.475 and its standard deviation is only 3.15. Then, we can consider here only the undirected and unweighted version of this network. Here, the nodes represent the countries and a link exists between two countries if one of them imports miscellaneous manufactures of metal (MMM) from the other. The undirected network is depicted in Fig. 8.

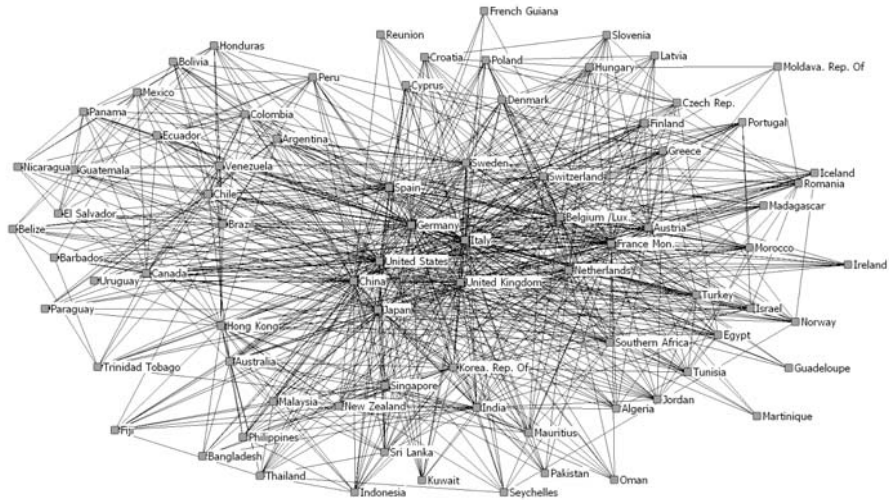

**Fig. 8.** Trade network of miscellaneous manufactures of metal (MMM). Nodes represent countries and a link exists between two countries if one exports MMM to the other.

We first calculated the communicability for every pair of countries in the dataset and then computed the $\Delta(G)$ matrix. Using this information we determined the number of cliques in the communicability graph, which correspond to the overlapped communities in the trade network. We found 27 communities with diverse degrees of overlapping. The overlapping matrix is illustrated in Fig. 9.

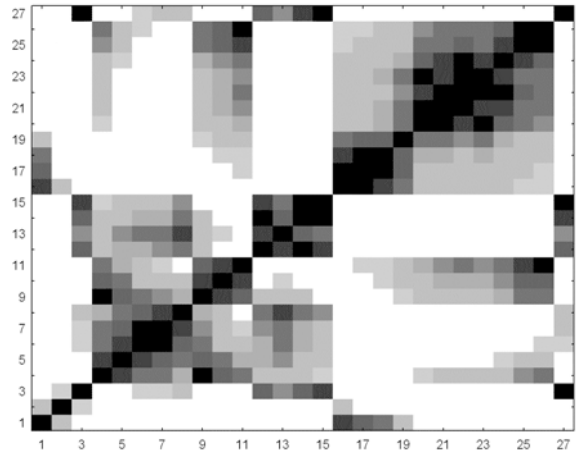

**Fig. 9.** Overlap matrix among the 27 communities found on the basis of the communicability for the trade network of MMM. Overlap increases from white to black in a scale from 0 to 1.

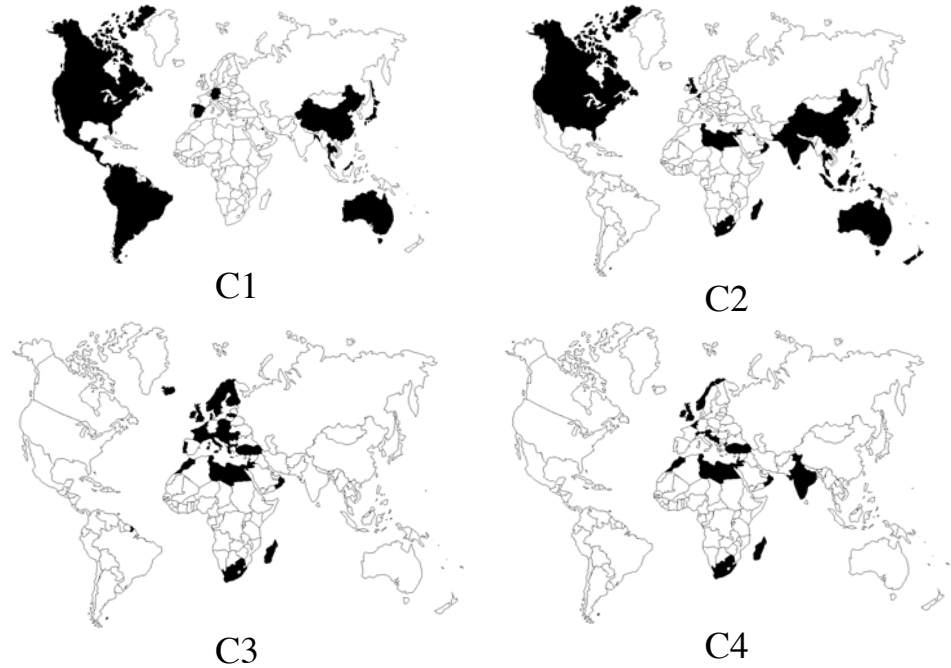

**Fig. 10.** Graphical representation of the countries forming four of the five communities having less than 50% of overlap in the trade network of miscellaneous manufactures of metal.

In order to obtain more valuable information for the analysis of this network we study the communities having less than 80% of overlap. In doing so, we have merged the communities with more than 80% of overlap following the procedure described in Section 3.2. Using this approach we reduced the number of communities to only 5, which will be analyzed in the following section.

## *6.1 Analysis of the communities*

The first community found is formed by 34 countries, 64.7% of which are located in America and 29.4% in Asia. Only two European countries, Germany and Spain, appear in this community. The second community is formed by 32 countries basically from Asia, as 62.5% of them are in this continent. The third community is formed by 39 countries, 61.54% of them are located in Europe. The fourth community is formed by 21 countries, one third of them are in Europe and the rest in Asia and Africa. The fifth community is the smallest one, which is formed by only 8 countries in Europe and the Caribbean (Italy, Greece, Spain, Sweden, Cyprus, French Guiana, Guadeloupe and Martinique). Four of these five communities are represented in Fig. 10.

A more detailed analysis of these communities reveals some interesting facts about the trade between these countries. The first community is dominated by the trade in and between America and Asia. It is not unexpected that Spain appears in this community instead of in the European one due to its historical relations with Latin-America. The reason why Germany is in this community is not clear. The second community reveals the trade between Anglo-Saxon countries (UK, USA, Canada and Australia) with Asian countries. The European community (C3 in Fig. 10) is mainly based on trade within them and with African and Middle-East countries. This community does include neither Spain nor Germany, which are instead included in the America-Asia community. This result reflects clearly the geopolitical nature of the trade clustering between countries. The fourth community reveals the trade between some European countries and some former African/Asian colonies of these European nations.

## *6.2 Trade under external stress*

We study here the effect of the external stress on the community structure of the trade network analyzed. The external stress is simulated here by the inverse temperature $\beta$. At the "normal" temperature $\beta = 1$ we have found 27 communities with diverse degrees of overlapping, as described above. These communities form a connected component. That is there is not any community which is isolated from the rest. As the temperature is increased ($\beta$ decreased) we start to observe the evolution of these communities under the effect of an external stressing factor. At $\beta = 0.017$ the first disconnection occurs in this trade network. A single country, Tunisia, is separated from the rest of the trading countries, leaving it isolated. At $\beta = 0.016$ a second country, Martinique, is disconnected, followed by Iceland, Spain and Honduras. In Fig. 11 we display the progress of this evolution as the temperature tends to infinity ($\beta \to 0$).

At the inverse temperature, $\beta = 0.008$, the network is split into six isolated pieces. The largest of these chunks is formed by the richest countries as USA, UK, Canada, Italy, France, Austria, Norway, Brazil, Argentina, Venezuela, Mexico, Norway, Chile, Switzerland and a few undeveloped countries like Uruguay, Egypt, Paraguay, Cyprus, Bangladesh, Colombia, Oman, Fiji, Nicaragua, Latvia and Malaysia. More details about this analysis will be published elsewhere [13].

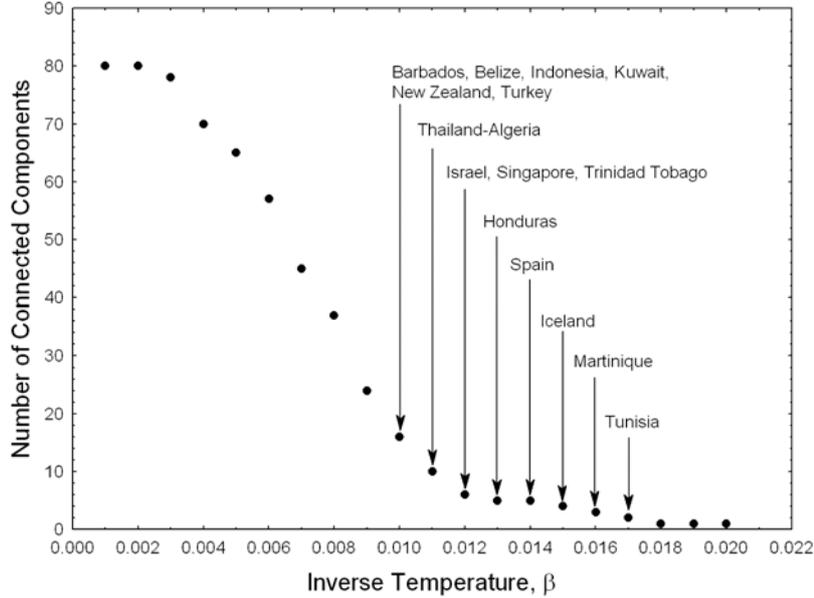

**Fig. 11.** Effect of external stress on the community structure of the trade network. The countries which are disconnected from the main trading network are represented at the temperature at which they disconnect.

## 7 Conclusions

We have introduced the concept of communicability in complex socio-economic networks. This concept allows a series of analysis from which we have selected the one related to the communities in socio-economic networks. The method for finding communities based on this concept permits the unambiguous identification of all communities in a network without the use of any external parameter. These communities display different level of overlapping, which can be managed *a posteriori* to merge communities according to specific necessities of the problem under analysis. The method is also unique in the sense that it permits to study the influence of external factors, like economic crisis and social agitation, by considering a network temperature.

We analyze here a trade network of miscellaneous metal manufactures between 80 countries. The communicability analysis revealed the existence of 27 communities with different level of overlapping. By considering only those communities with less than 80% of overlapping we have identified the existence of certain clusters that reflect the geopolitical relationships in the trade of such manufac-tures. Also important is the analysis of the resilience of these communities to external stresses. It has been shown that as the external "temperature" increases there are several countries which separate from the main trading network. These countries are the most vulnerable, the weakest links, in situations of critical external influences, like deep economical crisis. In closing, we hope that the communicability function and the method presented here for studying the identification and evolution of communities can add some valuable information to the study of complex socio-economic networks in the real-world.